\documentclass[aps,prd,reprint,showpacs,floatfix,a4pape, nofootinbib]{revtex4-1}

\usepackage[english]{babel}
\usepackage{graphicx}
\usepackage{array}
\usepackage{dcolumn}
\usepackage{bm}
\usepackage{amssymb}
\usepackage{amsmath}
\usepackage{amsfonts}
\usepackage{amsbsy}
\usepackage{hyperref}
\usepackage{breakurl}

\begin{document}

\title{Relativistic $\boldsymbol{\langle \sigma v_\texttt{rel}\rangle}$ in the calculation of 
relics abundances: a closer look }

\author{Mirco Cannoni}
\affiliation{Departamento de F\'isica Aplicada, Facultad de Ciencias
Experimentales, Universidad de Huelva, 21071 Huelva, Spain}

\begin{abstract}

In this paper we clarify the relation between the invariant relativistic relative velocity $V_{\texttt{r}}$, 
the M\o{}ller velocity $\bar{v}$, and the non-relativistic relative 
velocity $v_r$. Adopting $V_{\texttt{r}}$ as the true physical relative velocity for pair-collisions
in a non-degenerate relativistic gas, we show that in the frame co-moving with the gas
(\textit{i}) the thermally averaged  cross section times relative velocity 
$\langle \sigma v_\texttt{rel}\rangle$
that appears in the density evolution equation for thermal relics is reformulated
only in terms of  $V_{\texttt{r}}$ and $\mathcal{P}(V_{\texttt{r}})$ in a manifestly Lorentz invariant form;
(\textit{ii}) the frame-dependent issues of the standard formulation in terms of the M\o{}ller velocity,
as well as "superluminal" relative velocities, are not present in this formulation. 
Furthermore, considering  the annihilation of dark matter into a particle-antiparticle pair $f\bar{f}$, 
in the cases $m_f=0$, $m_f=m$ and $m_f \gg m$,
we find that the coefficients of the low velocity expansion of $\langle \sigma V_{\texttt{r}}\rangle$ 
admit an exact analytical representation
in terms of the Meijer $G$ functions that can be reduced to combinations of modified Bessel
functions of the second kind.

\end{abstract}

	
\date{November 17, 2013 }

\maketitle

\section{Introduction}
\label{Sec:intro}

In the Lee-Weinberg equation~\cite{LW} for the calculation of the  density of relic particles
in the expanding Universe
\begin{flalign}
\frac{d n}{dt}+3Hn=\langle \sigma v_{\text{rel}} \rangle  (n^2_{(0)}-n^2),
\label{LWequation}
\end{flalign}
a fundamental quantity is  $\langle \sigma v_{\text{rel}}\rangle$
where $\sigma$ is the total annihilation cross section and $v_{\text{rel}}$ the relative 
velocity of the annihilating particles. With $n_{(0)}$ we indicate the equilibrium distribution
of the number density and $H$ is the Hubble parameter.

If the relics at the freeze-out were non-relativistic, as in the case of the cold dark matter
of the standard cosmological model, then the system can be treated
as a non-relativistic classical gas. In this case, the meaning of thermal 
average $\langle...\rangle$ as well as "who" is $v_{\text{rel}}$ is clear.
In facts, the interaction rate 
\begin{flalign}
R=n_1 n_2 \sigma v_r,
\label{NR_rate}
\end{flalign}
contains the product of the cross section times  the magnitude of the non-relativistic relative velocity  
\begin{flalign}
{v}_{\textit{r}} = \rvert \boldsymbol{v}_1 - \boldsymbol{v}_2 \rvert.
\label{vRelNR}
\end{flalign}
The reaction rate is then averaged with the Maxwell distribution
$f_M (v)=(2/\pi)^{1/2}(m/T)^{3/2}v^2 \exp(-mv^2/(2T))$ for the absolute velocities
\begin{flalign}
\langle R \rangle =n_1 n_2 \int d^3\boldsymbol{v}_1 d^3\boldsymbol{v}_2 f_M ({v}_1) 
f_M({v}_2) \sigma 
{v}_{{r}}.
\label{NR_rate_ta}
\end{flalign}
By changing variables 
from the velocities $\boldsymbol{v}_1$, $\boldsymbol{v}_2$ to the velocity of 
the center of mass $\boldsymbol{v}_c $ and the relative velocity $\boldsymbol{v}_r$, 
one finds the standard expression for the thermal averaged rate,
\begin{flalign}
\langle R\rangle =n_1 n_2 \int_{0}^{\infty} dv_{{r}}  F_{\text{M}}(v_{{r}})\sigma v_{{r}} ,
\label{NR_rate_final}
\end{flalign}
where
\begin{flalign}
F_{{M}}(v_{{r}})=\sqrt{\frac{2}{\pi}}\left( \frac{\mu}{T}\right)^{3/2} %\left(\frac{m}{T}\right)^{3/2} 
{v}^2_{{r}}\, e^{-\frac{\mu}{T}\frac{v_{{r}}^2}{2}}
\label{Maxwell_vrel}
\end{flalign}
is the distribution of the relative velocity. Equation (\ref{Maxwell_vrel}) has the
same form of the Maxwell distribution for the absolute velocity but
with the reduced mass $\mu =m_1 m_2/(m_1 +m_2)$ in place of $m$
and $v_{{r}}$ in place of $v$. 
Considering the gas being composed by particles with mass $m$ such that $\mu=m/2$,
the thermally averaged cross section times the relative velocity is thus
\begin{flalign}
\langle \sigma  {v}_r\rangle&=\int_{0}^{\infty} d{v}_{{r}}F_{{M}}(v_{{r}}) \sigma {v}_{{r}}
%\crcr
=\frac{x^{3/2}}{2\sqrt{\pi}} \int_{0}^{\infty} d{v}_{{r}}
{{v}^2_{{r}}}
e^{-x \frac{{v}_{{r}}^2}{4}}   
\sigma {v}_{{r}},
\label{NRsigmav}
\end{flalign}
where we have introduced the standard thermal 
variable $x=m/T$. In this paper we use natural units with $\hbar=c=k_B =1$. 
The non-relativistic average (\ref{NRsigmav}) was used in the earlier calculation of the relic density,
see for example~\cite{GS}.

On the other hand,  the typical freeze-out temperature and masses of weakly interacting massive particles
are such that $x \sim m/T\sim 25$.
A rough estimate gives that the thermal velocity of the particles is $v~\sim \sqrt{3/x}\sim 0.35$,
thus relativistic corrections to (\ref{NRsigmav}) are expected.
 
Sredincki, Watkins and Olive~\cite{SWO} found the low-velocity and  large $x$ expansion 
of a relativistic formula based on the definition of $\langle\sigma v_{\text{rel}}\rangle$
given by Bernstein, Brown and Feinberg~\cite{BBF}.
In these papers the relative velocity is given by the expression~(\ref{vRelNR}).
Special relativity enters in the game by replacing the non-relativistic kinetic energy 
in the Maxwell-Boltzmann distribution with the relativistic $E=\sqrt{\boldsymbol{p}^2 +m^2}$
and using the standard definition of Lorentz invariant cross section.

Gondolo and Gelmini~\cite{GG} then re-derived the rate equation from the 
relativistic Boltzmann equation following the book~\cite{DeGroot}
and found that $v_{\text{rel}}$ in (\ref{LWequation}) is not the relative velocity~(\ref{vRelNR}),
but the so-called M\o{}ller velocity
\begin{flalign}
\bar{v}=\sqrt{(\boldsymbol{v}_1 - \boldsymbol{v}_2)^2 - (\boldsymbol{v}_1 \times \boldsymbol{v}_2)^2}.
\label{vMoller}
\end{flalign}
Starting from the general definition of thermal average
\begin{flalign}
\langle \sigma  \bar{v} \rangle=
\frac{\int d^3 \boldsymbol{p}_1 d^3 \boldsymbol{p}_2 e^{-E_1 / T} e^{-E_2 / T} 
\sigma \bar{v}}
{\int d^3 \boldsymbol{p}_1 e^{-E_1 / T} \int d^3 \boldsymbol{p}_2 e^{-E_2/ T}},
\label{Raverage_Def}
\end{flalign}
they showed that Eq.~(\ref{Raverage_Def}) reduces to the single-integral formula
\begin{flalign}
\langle \sigma  \bar{v} \rangle=\frac{1}{8 m^4 T K^2_2(x)}
\int_{4 m^2}^{\infty}
ds \sqrt{s} (s-4 m^2) K_1 \left(\frac{\sqrt{s}}{T}\right) \sigma.
\label{GGsigmav}
\end{flalign}
Here $s$ is the Mandelstam variable $s=(p_1 +p_2)^2$, $K_i$ are modified Bessel function
of the second kind. 
Equation~(\ref{GGsigmav}), and its extension to coahnnilation processes~\cite{EG}, was a step-forward
in the precise calculation of the relic density because, ones the annihilation cross section 
$\sigma(s)$ is known, the single integral can be calculated numerically
and does not necessitate any expansion or approximation and is used in public codes for relic density
calculation of dark matter.

Conceptually, anyway, formula (\ref{GGsigmav}) raises some questions:

(1) The integral on the right-hand side is manifestly Lorentz invariant 
but the M\o{}ller velocity, as much as the product $\sigma \bar{v}$, is not Lorentz invariant.
Thus the thermal average of the non-invariant quantity $\sigma \bar{v}$ turns out to be 
an invariant quantity.

(2) Comparing Eq.~(\ref{GGsigmav}) with Eq.~(\ref{NRsigmav}),
no velocity appears in the integral in the right-hand side of~(\ref{GGsigmav}).
Gondolo and Gelmini derived also a formula for $\langle \sigma \bar{v} \rangle$ 
that contains explicitly a velocity in the integral:
\begin{flalign}
\langle \sigma \bar{v} \rangle=\frac{2x}{K^2_2(x)}
\int_{0}^{\infty} \hspace{-0.3cm} d\varepsilon \sqrt{\varepsilon}\sqrt{1+2\varepsilon}
K_1 (2x\sqrt{1+\varepsilon}) \,\sigma \,(v_r)_{\text{lab}}.
\label{sv_GG_epsilon}
\end{flalign}
Here $\varepsilon =(s-4 m^2)/{4m^2}$ and $v_r$ are expressed in the lab frame, that is the rest frame of one particle.
In other words, the integral that defines the average of $\sigma \bar{v}$ 
in the form (\ref{sv_GG_epsilon}), implies that the co-moving frame coincides with the rest frame 
of one of the colliding particles.
As shown in \cite{GG}, one obtains a different result adopting the center of mass frame of the collision.
In this way a velocity has reappeared in the integral
but the Lorentz invariance of the integral is lost.

(3) As in the previous papers,  
the relative velocity~(\ref{vRelNR}) is considered as the natural expression
also in the relativistic framework. 
One further problem with $v_r$ and $\bar{v}$ is the fact that both can be larger than $c$ in the center of mass frame.

Given the general relevance of Eq.~(\ref{LWequation}) and Eq.~(\ref{GGsigmav}), 
the definition of the relativistic $\langle \sigma v_{\text{rel}}\rangle$ should be  free
of the exposed conceptual problems and should involve only Lorentz invariant quantities.
With this last statement we mean that, given the co-moving frame where the observer sees the gas 
at rest as a whole both $\sigma v_{\text{rel}}$ and the integral that gives average 
should be independent of frame where the kinematics of the collision is evaluated.

We thus try to answer the following questions:
\begin{enumerate} 
\item[-] 
Which is the correct relative velocity in special relativity that is Lorentz invariant 
and has values smaller or equal to $c$ in any inertial frame?
\item[-] 
Which is its probability density function in a relativistic classical gas? 
\item[-] 
Is it possible to define 
rate, flux and cross section in an Lorentz invariant way 
without using the non-invariant  M\o{}ller velocity?
\item[-] 
Is it possible to define $ \langle \sigma v_{\texttt{rel}} \rangle$ in the co-moving frame
as an integral over the probability distribution of the relative velocity in analogy with the non-relativistic case?
\end{enumerate}
We will see that the answering them we will solve the 3 raised problems.

The plan of the paper is the following.
In Section~\ref{Sec:2} we first review the concepts of relativistic relative velocity, M\o{}ller  
velocity and their relations with the definition of the invariant reaction rate and cross section.
Here we also clarify the paradox of superluminal relative velocities.

In Section~\ref{Sec:3}
we show that, actually, the velocity both in the left-hand side and right-hand side of (\ref{sv_GG_epsilon})
is $V_{\texttt{r}}$, the invariant relativistic relative velocity.
The relativistic thermal average in the co-moving frame 
$\langle \sigma  v_{\text{rel}} \rangle$  is $\langle \sigma V_{\texttt{r}} \rangle$ and the use 
of $\bar{v}$, $v_r$ and of any other reference frame can be avoided.

After that, in Section~\ref{Sec:4}   we  re-analyse the low-velocity expansion
of $\langle \sigma V_{\texttt{r}} \rangle$.
We find known and new expansions as well as 
exact relativistic expressions for the coefficients as a function of $x$
in some important cases.

The mathematics behind the  results of Section~\ref{Sec:2} and Section~\ref{Sec:4} is furnished 
by the relation between the generalized hypergeometric Meijer $G$ function and the modified Bessel functions. 
Being $G$ a special function that is not 
commonly encountered in particle and astro-particle physics,  
in Appendix~\ref{AppendixA} we  give a brief introduction 
and show its use in the calculation of the integrals. 

A summary of the main results is given in Section~\ref{summary}.

\section{Relative velocity,  M\o{}ller velocity and invariant rate
in Special Relativity }
\label{Sec:2}

In Special Relativity, the relative velocity between two massless particles and between a massless and 
a massive particle is $c$ in any frame, while the relative velocity between 
two massive particles is always smaller than $c$ in any frame. 
These requirements are not satisfied nor by ${v}_{\textit{r}}$ nor by $\bar{v}$.
For example in the center of mass frame of two colliding 
particles with mass $m$, $|\boldsymbol{v}_1|=|\boldsymbol{v}_2|=v$ and
$(\bar{v})_{\text{cm}}=({v_r})_{\text{cm}}=2v$, 
thus, for $v>1/2$, $\bar{v}$ and $v_r$ assume non-physical values larger than $c$.

On the other hand, a relative velocity compatible 
with the principles of Special Relativity is well known~\cite{Landau2,Cercignani}
and is given by
\begin{flalign}
V_{\texttt{r}}=
\frac{\sqrt{(\boldsymbol{v}_1 - \boldsymbol{v}_2)^2 - (\boldsymbol{v}_1 
\times 
\boldsymbol{v}_2)^2}}
{1-\boldsymbol{v}_1 \cdot \boldsymbol{v}_2}.
\label{vRelRel}
\end{flalign} 
This expression is symmetric in the two velocities and is valid in any frame.
When one particle (or both) is  massless, then $|\boldsymbol{v}_i|=1$ and also $V_{\texttt{r}}=1$. 
Considering the example above, in the center of mass frame we have
$(V_{\texttt{r}})_{\text{cm}}={2v}/{(1+v^2)}$,
thus, differently from $\bar{v}$ and  $v_r$, $V_{\texttt{r}}$ is always smaller or equal to velocity of light.
In the non-relativistic limit $V_{\texttt{r}}$, as much as $\bar{v}$, reduces to (\ref{vRelNR}).
The expression of $v_\text{r}$ in terms of the four-momentum $p_{1,2}$ is 
\begin{flalign}
 V_{\texttt{r}} =\frac{\sqrt{(p_1 \cdot p_2)^2 -m^2_1 m^2_2}}{p_1 \cdot p_2},
\label{vRel_p1p2}
\end{flalign} 
that manifests the  Lorentz invariance.

Although $\langle \sigma v_{\text{rel}}\rangle $ is often called simply thermally 
averaged cross section, actually the quantity is the thermal average of the 
interaction (or annihilation) rate per unit density. We thus now reformulate
the non-relativistic expression (\ref{NR_rate})
in a Lorentz invariant way using the invariant relative velocity $V_{\texttt{r}}$.
The expression of  the invariant rate valid in any frame is~\cite{Landau2,Cercignani}
\begin{flalign}
\mathcal{R}= n_1 n_2 \frac{p_1\cdot p_2}{E_1 E_2} \sigma V_{\texttt{r}}.
\label{Lorentz_inv_rate}
\end{flalign}
The factor 
\begin{flalign}
\frac{p_1\cdot p_2}{E_1 E_2}=\frac{\gamma_\texttt{r}}{\gamma_1 \gamma_2}=1-\boldsymbol{v}_1 \cdot \boldsymbol{v}_2 
\label{factor}
\end{flalign} 
accounts for the Lorentz contraction of the volumes of number densities in a generic frame
and assures the Lorentz invariance of the product $n_1 n_2 {p_1\cdot p_2}/{(E_1 E_2)}$. In (\ref{factor}) $\gamma_{1,2}$ and 
$\gamma_\texttt{r}$ are the Lorentz factors $\gamma=1/\sqrt{1-v^2}$ associated to the corresponding velocities. 

From the definition of the invariant rate it follows the definition of the invariant cross section:  
$\sigma={\mathcal{R}}/{F}={1}/{F} \int \overline{|\mathcal{M}|^2} \text{d}\Phi{(f)}$,
where $\overline{|\mathcal{M}|^2}$ is the squared matrix element summed over final spins and averaged over the 
initial spins, ${d}\Phi{(f)}$ is the usual Lorentz invariant phase space for the final state particles,
and $F$ is the invariant flux
\begin{flalign}
F=n_1 n_2 \frac{p_1\cdot p_2}{E_1 E_2} V_{\texttt{r}}.
\label{invF} 
\end{flalign}
If we normalize the one-particle states to $2E$, that is the number of 
particles per unit volume are $2E$,
the number density is $n=2E$, then the invariant flux (\ref{invF}) becomes
\begin{flalign}
F=4( p_1\cdot p_2) V_{\texttt{r}} =4\sqrt{(p_1\cdot p_2)^2 -m^2_1 m^2_2},
\label{F_vrel}
\end{flalign}
and the standard formula  for the invariant cross section is obtained.

An equivalent formulation is obtained by introducing the  M\o{}ller velocity
$\bar{v}$.
In facts, using (\ref{vMoller}), (\ref{vRelRel}) and (\ref{vRel_p1p2}) we have 
\begin{flalign}
\bar{v}={(1-\boldsymbol{v}_1 \cdot \boldsymbol{v}_2)}{V_{\texttt{r}}}=
\frac{p_1 \cdot p_2}{E_1 E_2}V_{\texttt{r}} =
\frac{\sqrt{(p_1 \cdot p_2)^2 -m^2_1 m^2_2}}{E_1 E_2}.
\label{Moll_from_Rel}
\end{flalign} 
The invariant rate (\ref{Lorentz_inv_rate}) is
\begin{flalign}
\mathcal{R}=n_1 n_2 \sigma \bar{v},
\label{Lorentz_inv_rate_vmoller}
\end{flalign}
and the invariant flux, with same normalization of the densities, 
$F=4 E_1 E_2\bar{v}
=4\sqrt{(p_1\cdot p_2)^2 -m^2_1 m^2_2}$.

The only reason to introduce the M\o{}ller velocity
is to write the Lorentz invariant rate (\ref{Lorentz_inv_rate}) in the form  
(\ref{Lorentz_inv_rate_vmoller}) that is similar to the non relativistic 
expression (\ref{NR_rate}). 
Anyway, while in the latter each factor is Galileo invariant, 
relativistically the products $n_1 n_2 ({p_1\cdot p_2}/{E_1 E_2})$ and $n_1 n_2 \bar{v}$ 
are Lorentz invariant.
This redefinition is the reason why the collision term of the 
relativistic Boltzmann equation has the same form as the one in the non-relativistic 
equation with $v_r$ replaced by $\bar{v}$~\cite{DeGroot, Cercignani}.

It should be clear from the previous discussion that neither $v_r$ nor $\bar{v}$ are  the relative velocity 
in special relativity, contrary to what is often claimed in literature and textbooks. 
In textbooks, when defining the invariant cross section, the flux is usually defined
in a frame where the velocities are collinear, say the lab or cm frame,
\begin{flalign}
F^{\text{coll}}=n_1 n_2 |\boldsymbol{v}_1 -\boldsymbol{v}_2|,
\label{F_coll}
\end{flalign}
in analogy with the non relativistic case $n_1 n_2 v_r$, 
and then it is rewritten in the invariant form using
$|\boldsymbol{v}_1 -\boldsymbol{v}_2|=\sqrt{(p_1\cdot p_2)^2 -m^2_1 m^2_2}/E_1 E_2$
and $n_1 n_2 =4E_1 E_2$.
It is the formal equivalence of $v_r$ with the quantity $ |\boldsymbol{v}_1 -\boldsymbol{v}_2|$ 
in the definition of the 
flux that is probably at the origin of the confusion about "who" is the relativistic relative velocity and 
the paradox that the relative velocity can have values larger than the velocity of the light.

Actually, although $ |\boldsymbol{v}_1 -\boldsymbol{v}_2|$ is mathematically equivalent 
to non-relativistic relative velocity $v_{r}$, conceptually the two quantity have nothing to do with each other.
In the relativistic framework 
\begin{flalign}
|\boldsymbol{v}_1 -\boldsymbol{v}_2|=\left(\frac{p_1\cdot p_2}{E_1 E_2} V_{\texttt{r}}\right)^{\text{coll}}
=(\bar{v})^{\text{coll}}.
\end{flalign}
From this point of view the fact that $ |\boldsymbol{v}_1 -\boldsymbol{v}_2|$ assumes in the center of mass  frame 
values larger than $c$ is not 
a problem because this quantity  is not the  
relative velocity. For example, consider the scattering of
two massless particles as seen in the cm frame. The relative velocity is $V_{\texttt{r}}=1$, but
$p_1\cdot p_2/{(E_1 E_2)}=2$, thus $|\boldsymbol{v}_1 -\boldsymbol{v}_2|=\bar{v}=2$: the relative velocity
is never larger than 1, $\bar{v}$ can be larger than 1 because is not a physical velocity.

We thus addressed the problem (3) of the Introduction.

\section{Relativistic $\boldsymbol{\langle \sigma V_{\texttt{r}}\rangle}$}
\label{Sec:3}

Under the hypothesis that the system can be treated as a non-degenerate relativistic gas
in equilibrium~\cite{BBF,GG}, the normalized  momentum distribution in the comoving frame is given by
the J\"{u}ttner~\footnote{We conform to the practice, common to other fields
of Physics where relativistic thermodynamics is used~\cite{DeGroot,Cercignani,csernai}, 
to attribute the distribution~(\ref{F_J}) to J\"{u}ttner
that found it in Ref.~\cite{juttner}.} distribution
\begin{equation}
f_{J}(\boldsymbol{p})= \frac{1}{4\pi m^2 T K_2 (x)} e^{-\frac{\sqrt{\boldsymbol{p}^2+m^2}}{T}},
\label{F_J}
\end{equation}
Averaging the rate (\ref{Lorentz_inv_rate}) with the J\"{u}ttner distribution (\ref{F_J}), the relativistic analogous 
of Eq.~(\ref{NR_rate}) hence is 
\begin{equation}
\langle \mathcal{R}\rangle=n_1 n_2 \int \frac{d^3 \boldsymbol{p}_1}{E_1} \frac{d^3 \boldsymbol{p}_2}{E_2} 
 f_{J}(\boldsymbol{p}_1) f_{J}(\boldsymbol{p}_2)\,
(p_1\cdot p_2)  \sigma   v_\texttt{r}.
\label{Rel_averaged_rate}
\end{equation}
In Ref.~\cite{Cannoni1} we have shown that
\begin{flalign}
\int \frac{d^3 \boldsymbol{p}_1}{E_1}  \frac{d^3 \boldsymbol{p}_2}{E_2}
{p_1\cdot p_2} f_{J}(\boldsymbol{p}_1) f_{J}(\boldsymbol{p}_2) 
\equiv \int^{1}_{0} dV_{\texttt{r}} \mathcal{P}_{\texttt{r}}(V_{\texttt{r}}) =1,
\nonumber
\end{flalign}
where $\mathcal{P}_{\texttt{r}}(V_{\texttt{r}})$ is the probability density distribution of $V_{\texttt{r}}$:
\begin{equation}
\begin{array}{l}
\mathcal{P}_{\texttt{r}}(V_{\texttt{r}})=
\frac{X}{\sqrt{2} \prod_{i} K_2 (x_i) }
 \frac{\gamma^3_{_{\texttt{r}}} (\gamma^2_{_{\texttt{r}}} -1)}{\sqrt{\gamma_{{\texttt{r}}} +\varrho}}
K_1 (\sqrt{2}X\sqrt{\gamma_{{\texttt{r}}} +\varrho}).
\end{array}
\label{P_v}
\end{equation}
with the abbreviations
\begin{equation}
X=\sqrt{x_1 x_2},\;\;\varrho=\frac{m^2_1 +m^2_2}{2m_1 m_2}=\frac{x^2_1 +x^2_2}{2x_1 x_2}.
\label{X_ro}
\end{equation}

In the following we adopt the symbol $\langle ...\rangle_\mathcal{P}$ to indicate that the average 
of a certain quantity is obtained integrating  over  $\mathcal{P}(V_{\texttt{r}})$
in agreement with the general methods of statistical mechanics:
$V_{\texttt{r}}$ is the physical Lorentz invariant quantity,
admits a normalized probability density function $\mathcal{P}(V_{\texttt{r}})$. 
Given any  $f(V_{\texttt{r}})$ its average is obtained integrating it over  $\mathcal{P}(V_{\texttt{r}})$.
This is in complete analogy with the non-relativistic average (\ref{NRsigmav})
where the relative velocity is $v_r$, Eq.~(\ref{vRelNR}), and the probability density function is given by the 
Maxwell distribution $F_M (v_r)$,~Eq.~(\ref{Maxwell_vrel}).

Having clarified the concept of relativistic relative velocity and having found its
probability distribution, it is clear that the thermal average of  
$\sigma V_{\texttt{r}}$ is
\begin{flalign}
\langle \sigma V_{\texttt{r}} \rangle_\mathcal{P}=
\int^{1}_{0} dV_{\texttt{r}} \mathcal{P}_{\texttt{r}}(V_{\texttt{r}})\sigma V_{\texttt{r}}.
\label{sv_vrel_P}
\end{flalign}
We can write the average in terms of the more 
practical variables $\gamma_{\texttt{r}}$ and $s$:
\begin{flalign}
&\langle \sigma V_{\texttt{r}} \rangle_\mathcal{P} =
\frac{X}{\sqrt{2} \prod_{i} K_2 (x_i)}\crcr
& \times \int^{\infty}_{1} d\gamma_{\texttt{r}} 
\gamma_{_{\texttt{r}}} \sqrt{\frac{\gamma^2_{_{\texttt{r}}}-1}{\gamma_{_{\texttt{r}}}+\varrho}}
K_1 (\sqrt{2}X\sqrt{\gamma_{_{\texttt{r}}} +\varrho}) \sigma V_{\texttt{r}}\nonumber
\\
&= \frac{1}{4 T \prod_{i} m^2_i K_2 (x_i)}\crcr
& \times  \int^{\infty}_{M^2}
ds  [s-(m^2_1 +m^2_2)] p' K_1 (\sqrt{s}/T) \sigma V_{\texttt{r}},
\label{sigmav_s} 
\end{flalign}
where $p'$ is 
\begin{equation}
p'= \frac{\sqrt{s-(m_1 + m_2)^2} \sqrt{s-(m_1 - m_2)^2} }    {2\sqrt{s}}.
\end{equation}
and  $M=m_1 +m_2$.

The relative velocity in the integrand can be eliminated using the relations
$V_{\texttt{r}}={\sqrt{\gamma^2_{\texttt{r}}-1}}/{\gamma_{\texttt{r}}}$
and, from Eq.~(\ref{vRel_p1p2}),  
$s=(m_1 - m_2)^2 +2 m_1 m_2 (1+\gamma_{\texttt{r}})$ that give
\begin{flalign}
V_{\texttt{r}}
=\frac{\sqrt{[s-(m_1 + m_2)^2][s-(m_1 - m_2)^2] }}    {s-(m^2_1+m^2_2)}.
\label{vrel_s}
\end{flalign}
The formulas as integrals over the cross section are: 
\begin{flalign}
&\langle \sigma V_{\texttt{r}} \rangle_\mathcal{P}
=\frac{X}{\sqrt{2} \prod_{i} K_2 (x_i)} \crcr
& \times 
\int^{\infty}_{1} d\gamma_{\texttt{r}} 
\frac{\gamma^2_{_{\texttt{r}}}-1}{\sqrt{\gamma_{_{\texttt{r}}}+\varrho}}
K_1 (\sqrt{2}X\sqrt{\gamma_{_{\texttt{r}}} +\varrho}) \sigma\nonumber
\\
& = \frac{1}{8 T \prod_{i} m^2_i K_2 (x_i)}
\int^{\infty}_{M^2} ds \frac{[s-(m_1 -m_2)^2]}{\sqrt{s}}\crcr
&\times [s-(m_1 + m_2)^2] K_1 ({\sqrt{s}}/{T}) \sigma.
\label{int_s}
\end{flalign}
We remark that the formulas in (\ref{sigmav_s}) and (\ref{int_s})
as a function of $\gamma_\texttt{r}$ are more general than the formulas
as a function of $s$. The former are valid also for different temperatures. In facts the temperature
enters in the formula only through the scaling variable $x_i =m_i /T_i$, hence even if $m_1 \neq m_2$ and $T_1 \neq T_2$
the expression does not change because $X$ and $\varrho$ maintain the same dependence on $x_1$ and $x_2$ given by 
Eq.~(\ref{X_ro}), as it is easy to verify.

We now compare these formulas with those  discussed in the Introduction 
and find the answer to conceptual questions (1) and (2) raised there.

(1) Setting $m_1=m_2=m$ in Eq.~(\ref{int_s})
we find the integral on the right-hand side of Eq.~(\ref{GGsigmav}).
The integral in the numerator of Eq.~(\ref{Raverage_Def}) is not invariant because
both the phase-space elements $d^3 \boldsymbol{p}_i$ and $\bar{v}$ are not. However, we have seen that 
$\bar{v}=(p_1 \cdot p_2)/(E_1 E_2) V_{\texttt{r}}$, thus ones we explicit this relation, the integral is the same as the one 
in Eq.~(\ref{Rel_averaged_rate}).
Equation (\ref{GGsigmav}) gives the correct result because $V_{\texttt{r}}$ and  $\mathcal{P}(V_{\texttt{r}})$ are present in the integral but are hidden by $\bar{v}$:
actually, one is calculating  $\langle \sigma V_{\texttt{r}} \rangle_{\mathcal{P}}$.
Put in other words, $\langle \sigma \bar{v}\rangle$ is only a different notation for $\langle \sigma V_{\texttt{r}} \rangle_{\mathcal{P}}$ as much as Eq.~(\ref{Lorentz_inv_rate_vmoller}) is a different way of writing 
Eq.~(\ref{Lorentz_inv_rate}).

2) Setting $m_1=m_2=m$ and  using the variable $\varepsilon=(s-4m^2)/4m^2$ 
in Eq.~(\ref{sigmav_s}) we get 
\begin{flalign}
\langle \sigma V_{\texttt{r}} \rangle_\mathcal{P}=\frac{2x}{K^2_2(x)}
\int_{0}^{\infty} d\varepsilon \sqrt{\varepsilon} \sqrt{1+2\varepsilon}
K_1 (2x\sqrt{1+\varepsilon}) \sigma V_{\texttt{r}}.
\nonumber
\end{flalign}
Comparing with Eq.~(\ref{sv_GG_epsilon}), we see that 
(\ref{sv_GG_epsilon}) is nothing but 
$\langle \sigma  \bar{v} \rangle_\mathcal{P}$
with the integral evaluated in the rest frame of one particle:
in facts, only in this frame $(\bar{v})_{\text{lab}}=(v_r)_{\text{lab}}=(V_{\texttt{r}})_{\text{lab}}$ 
and $\langle \sigma  \bar{v} \rangle_\mathcal{P}$
coincides with $\langle \sigma V_{\texttt{r}} \rangle_\mathcal{P}$. 
Strictly speaking, the averages $\langle \bar{v} \rangle_\mathcal{P}$ and
$\langle \sigma  \bar{v} \rangle_\mathcal{P}$ have not a precise physical meaning.
To calculate the average $\langle \bar{v} \rangle_\mathcal{P} =\int dV_{\texttt{r}} \mathcal{P}(V_{\texttt{r}}) \bar{v}$,
the co-moving frame is not enough: one has to specify $\bar{v}$ in the rest frame of one particle or in the center of mass frame. In the former case coincides with 
$\langle V_{\texttt{r}}\rangle_\mathcal{P}$, but will have a different form and take different values in the center of mass frame.

We stress again that ones it is recognized that $V_{\texttt{r}}$, Eqs.~(\ref{vRelRel}), (\ref{vRel_p1p2}), (\ref{vrel_s}),
\textit{is the relative velocity} in the relativistic framework and (\ref{F_vrel}) is the invariant flux, then
the invariant cross section follows automatically, a probability density function for $V_{\texttt{r}}$ exists
and the average value of the cross section times the relative velocity is obtained integrating over the probability density.
The only reference frame is the co-moving frame where the observer measure the velocities of the colliding particles.
All the problems connected with the non-invariance of the M\o{}ller velocity and the superluminal values in the center of mass frame disappear. 

All the conceptual problems are eliminated considering $\bar{v}$ just a short-hand notation for 
$p_1\cdot p_2/(E_1 E_2) V_{\texttt{r}} $ and not as a velocity to work with in formulating
relativistic concepts.

\section{Low velocity expansion }
\label{Sec:4}

We skip here the subscript $\mathcal{P}$ to lighten the notation.
Let us consider the typical annihilation of two dark matter particles with mass $m$
into particle-antiparticle two-body final states with mass $m_{{f}}$, $X X\to f\bar{f}$.
We express the total cross section as the integral of the differential cross section in 
center of mass frame that we write as:
\begin{flalign}
\sigma=\frac{2m^2}{ s} \frac{\sqrt{s-4m^2_{{f}}}}{\sqrt{s-4m^2}}
\sigma_0,\;\;\sigma_0 =\frac{1}{2m^2}\frac{1}{64 \pi^2} \int \overline{\lvert\mathcal{M}\lvert^2} d\Omega_{\text{cm}}.
\nonumber
%\label{sigma_con_s0}
\end{flalign}
For what follows it is convenient to use the dimensionless variable
$y={s}/(2m)^2=(\gamma_{\texttt{r}}+1)/2$. Changing variable in (\ref{int_s}) 
with $m_1 =m_2 =m$ we obtain 
\begin{flalign}
\langle \sigma V_{\texttt{r}} \rangle=\frac{2x}{K^2_2(x)}
\int_{y_{0}}^{\infty} \frac{dy}{\sqrt{y}} \sqrt{y-1}\sqrt{y-\rho} K_1 (2x\sqrt{y}) \sigma_0(y),
\nonumber
%\label{integral_y}
\end{flalign}
where, setting $\rho=m^2_f /m^2$, the inferior limit is $y_{0}=1$ if $m \geq m_f$, and $y_0 =\rho$ if $m<m_f$.

Except the case when  there are resonant propagators with small width of the exchanged particles,
$\sigma_0$ is a well behaved smooth function. For cold dark matter $y-1$ is a small quantity, thus
$\sigma_0$ can be expanded in powers of $y-1$,
\begin{flalign}
\sigma_0(y)=\sum\limits_{n=0}^{\infty} \frac{1}{n!} \sigma^{(n)}_0 (y -1)^n.
\end{flalign}
where the derivatives $\sigma^{(n)}_0$ are evaluated at $y=1$.
We can write the low-velocity expansion as
\begin{flalign}
\langle \sigma V_{\texttt{r}} \rangle= %\frac{2x}{K^2_2(x)}
\sum\limits_{n=0}^{\infty} \frac{1}{n!} \sigma_0^{(n)}
\mathcal{K}^\rho_n (x)
\label{expansion}
\end{flalign}
where the coefficients are given by
\begin{flalign}
\mathcal{K}^\rho_n (x)&=\frac{2x}{K^2_2(x)} I^\rho_n (x),
\label{K_n_rho}\\
I^\rho_n (x)&=
\int_{y_0}^{\infty}\frac{dy}{\sqrt{y}}
(y-1)^{(1/2 +n)}\sqrt{y -\rho}  K_1 (2x\sqrt{y}).
\label{I_n}
\end{flalign}
For $\rho=0$, $\rho=1$ and $\rho\gg 1$,  the integral (\ref{I_n})
can be reduced to the known integral~\cite{GR} and the properties
of the generalized hypergeometric 
Meijer's $G$ function~\cite{GR} discussed in the Appendix.
We study them separately.

\subsection{$\boldsymbol{\rho=0 }$}

When the mass of the annihilation products is
much smaller the mass of the dark matter particle we have $\rho\ll 1$
and $\rho=0$ gives a good approximation.  
In this case
\begin{equation}
I^{\rho=0}_n (x)=
\int_{1}^{\infty}
dy (y-1)^{1/2 +n}   K_1 (2x\sqrt{y})
\end{equation}
corresponds to~(\ref{I_GR}) with $\lambda=0$, $\mu=3/2+n$ and $\nu=1$. Hence
\begin{equation}
I^{\rho=0}_n (x)
=\frac{1}{2} \Gamma \left(n+\frac{3}{2}\right) 
G_{1,3}^{3,0}
\begin{pmatrix}
x^2 \Biggr\rvert
\begin{array}{c}
 0 \\
-n-\frac{3}{2}, \frac{1}{2}, -\frac{1}{2}
\end{array}
\end{pmatrix}.
\end{equation}
The coefficients (\ref{K_n_rho}) of
the low-velocity expansion~(\ref{expansion}) are 
\begin{flalign}
\mathcal{K}^{\rho=0}_n (x)=
\Gamma(n+\frac{3}{2}) 
\frac{x}{ K^2_2(x)}
G_{1,3}^{3,0}
\begin{pmatrix}
x^2 \Biggr\rvert
\begin{array}{c}
 0 \\
 -\frac{1}{2},\frac{1}{2},-n-\frac{3}{2}
\end{array}
\end{pmatrix}.
\label{expansion_rho=0}
\end{flalign}

The expansion in powers of $x^{-1}$ should 
give the standard result of Srednicki, Watkins and Olive~\cite{SWO}.
To study the large $x$ behavior, we remind that for $x\gg1$
the Bessel functions are approximated by~\cite{GR}
\begin{equation}
K_n (z)={e^{-z}}\sqrt{\frac{\pi}{2z}}(1+\frac{4n^2-1}{8 z}+\frac{16 n^4 -40 n^2+9}{128 z^2}+...),
\label{Bessel_expansion}
\end{equation} 
while  the asymptotic expansion of the $G$ function for $x\gg 1$ for some $n$~\cite{GR,Mathematica}: 
\begin{equation}
\begin{array}{l}
G_{1,3}^{3,0}\left(x^2\Bigg {\rvert}
\begin{array}{c}{}
0 \\
-\frac{1}{2},\frac{1}{2},-n-\frac{3}{2}
\end{array}
\right)=\sqrt{\pi}e^{-2x}    \\
\\
\qquad\qquad   \times\left\{
\begin{array} {lc}
 \left(\frac{1}{x^2}+\frac{3}{4}\frac{1}{x^3}-\frac{3}{32}
\frac{1}{x^4}+\mathcal{O}(x^{-5})\right), & n=0\\
\\
 \left(\frac{1}{x^3}+\frac{7}{4}\frac{1}{x^4}+\mathcal{O}(x^{-5})\right), & n=1\\
\\
\left(\frac{1}{x^4}+\mathcal{O}(x^{-5})\right), & n=2\\
\\
\left(\frac{1}{x^5}+\mathcal{O}(x^{-6})\right) & n=3.
\end{array}\right.
\end{array}
\label{Gexpansion}
\end{equation}
For a given $n$, the polynomial starts with the power $x^{-(n+2)}$
and continues with increasing powers $x^{-(n+3)}$,  $x^{-(n+4)}$ and so on,
while
\begin{equation}
\frac{x}{K^2_2(x)}={e^{2 x}} \frac{2}{\pi} 
\left(x^2-\frac{15}{4} x+\frac{285}{32} 
+\mathcal{O}(x^{-1})\right),
\label{xK22expansion}
\end{equation}
starts with $x^2$ and continues decreasing the exponent by steps of 1. 
The expansion of the product $x/K^2_2 (x) G_{1,3}^{3,0}$, for each $n$, thus starts with $x^{-n}$.
If we want the thermal averaged cross section up to the order $x^{-3}$, we have only to consider the terms 
with $n$ from 0 to 3:
\begin{flalign}
\langle \sigma {v}_{\texttt{r}} &\rangle=
\sigma_0^{(0)}\mathcal{K}_0 (x)_{(\mathcal{O}(x^{-4}))}+\sigma_0^{(1)}\mathcal{K} _1 (x)_{(\mathcal{O}(x^{-4}))}\cr
&+\frac{1}{2}\sigma_0^{(2)}\mathcal{K}_2 (x)_{(\mathcal{O}(x^{-4}))}+\frac{1}{6}\sigma_0^{(3)}\mathcal{K}_3 
(x)_{(\mathcal{O}(x^{-4}))}.
\label{expansion3}
\end{flalign}
Using in (\ref{expansion3}) the expansions (\ref{Gexpansion}), (\ref{xK22expansion}) and the coefficients
\begin{equation}
\Gamma\left(\frac{3}{2}+n\right)=\frac{(2n+1)!!}{2^{n+1}}\sqrt{\pi},
\end{equation}
we finally find up to $\mathcal{O}({x^{-4}})$, 
\begin{flalign}
\langle \sigma {v}_{\texttt{r}} &\rangle=
\sigma_0^{(0)}
+\left(-3\sigma_0^{(0)}+\frac{3}{2}\sigma_0^{(1)}\right)\frac{1}{x}\cr
&+\left(6\sigma_0^{(0)}-3\sigma_0^{(1)}+\frac{15}{8} \sigma_0^{(2)}\right)\frac{1}{x^2}\nonumber\\
&-\frac{5}{16} \left(30\sigma_0^{(0)}-15\sigma_0^{(1)}+3\sigma_0^{(2)}-7\sigma_0^{(3)}\right)\frac{1}{x^3},
\end{flalign}
that is the expected result~\cite{SWO}.
This new derivation based on the property of the $G$ function shows clearly why the coefficients of the powers 
$x^{-n}$ are linear combinations of the derivatives $\sigma_0^{(n)}$ starting from $n=0$ up the order of the corresponding power of $x$.   

\subsection{$\boldsymbol{\rho=1}$}

The integral~(\ref{I_n}) is
\begin{eqnarray}
I^{\rho=1}_n (x)=
\int_{1}^{\infty}
dy y^{-1/2}(y-1)^{1 +n}  K_1 (2x\sqrt{y}) ,
\end{eqnarray}
that reduces to (\ref{I_GR}) with $\lambda=-1/2$, $\mu=2+n$ and $\nu=1$:
\begin{flalign}
I^{\rho=1}_n (x)&=\Gamma(2+n)\frac{x}{4} 
G_{1,3}^{3,0}
\begin{pmatrix}
x^2 \Biggr\rvert
\begin{array}{c}
 0 \\
-n-2, 0, -1
\end{array}
\end{pmatrix}
\nonumber\\
&=\Gamma (2+n) \frac{1}{x^{(2+n)}}  K_{n+1}(2 x).
\end{flalign}
The last equality is proved in Appendix~\ref{A1}.
The coefficients (\ref{K_n_rho}) of
the low-velocity expansion~(\ref{expansion}) are: 
\begin{flalign}
\mathcal{K}^{\rho=1}_n (x)=
\Gamma (n+2)\frac{2}{x^{n+1}} \frac{ K_{n+1}(2 x)}{K^2_2 (x)}.
\label{expansion_rho=1}
\end{flalign}
Using (\ref{Bessel_expansion}) and (\ref{xK22expansion}) we find the large $x$ expansion:
\begin{flalign}
\langle \sigma {v}_{\texttt{r}}\rangle_{\rho=1} 
&=\frac{2x^{-1/2}}{\sqrt{\pi}} [\sigma_0^{(0)}+(-\frac{57}{16}\sigma_0^{(0)}+ 2 
\sigma_0^{(1)})\frac{1}{x}\crcr
&+ 3( \frac{1395}{512} \sigma_0^{(0)} -\frac{960}{512} \sigma_0^{(1)} + \sigma_0^{(2)} ) \frac{1}{x^2}+...].
\end{flalign}
This expansion never appeared before.

\subsection{$\boldsymbol{\rho\gg 1}$}

Neglecting the unity in $(y-1)^{n+1/2}$ and with $\rho$ as lower limit 
of integration  we find after changing variable to $y/\rho$,
\begin{flalign}
&I^{(\rho \gg 1)}_n (x)=
\int_{\rho}^{\infty}
dy y^n \sqrt{y -\rho} K_1 (2x\sqrt{y}) \crcr
&=\rho^{3/2+n}\int_{1}^{\infty} dy y^n \sqrt{y-1} K_1(2x\sqrt{\rho}\sqrt{y}),
\end{flalign}
that reduces to (\ref{I_GR}) with $\lambda=n$, $\mu=3/2$ and $\nu=1$:
\begin{flalign}
I^{(\rho \gg 1)}_n (x)=\frac{\sqrt{\pi }}{4} \frac{\rho ^{3/2}}{x^{2 n}} 
G_{1,3}^{3,0}\left(\rho x^2  \Biggr\rvert 
\begin{array}{c}
 0 \\
-\frac{3}{2}, \frac{1}{2}+n,-\frac{1}{2}+n
\end{array}
\right).
\end{flalign}
The coefficients (\ref{K_n_rho}) of
the low-velocity expansion~(\ref{expansion}) are:
\begin{flalign}
\mathcal{K}^{\rho\gg 1}_n (x)=
\frac{\sqrt{\pi} \rho ^{3/2}x}{2 x^{2 n} K^2_2 (x)}
G_{1,3}^{3,0}\left(\rho x^2   \Biggr\rvert 
\begin{array}{c}
0 \\
-\frac{3}{2},\frac{1}{2}+n,-\frac{1}{2}+n
\end{array}\right).
\label{expansion_rho>>1}
\end{flalign}
For the expansion at large $x$ we find
\begin{equation}
\langle \sigma {v}_{\texttt{r}} \rangle_{\rho\gg 1} \approx
e^{-2x(\sqrt{\rho} -1)}[\sqrt{\rho}(\sigma_0^{(0)} + \rho \sigma_0^{(1)}) +\mathcal{O}(x^{-1})],
\end{equation}
that show the well-known suppression for heavy masses~\cite{GS,GG}.
Expansions~(\ref{expansion_rho=1}) and (\ref{expansion_rho>>1}) correspond to the cases "at threshold" and 
"above threshold", respectively first discussed in Refs.~\cite{GS,GG} in the non-relativistic case.
\begin{figure}[t!]
\includegraphics*[scale=.6]{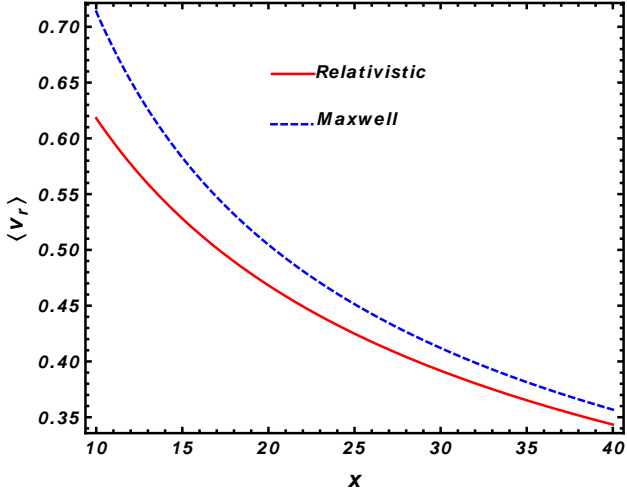}
\caption{Mean relative velocity as a function of $x=m/T$. The red line is the relativistic
value in Eq.~(\ref{sigma_cost}), see also~(\ref{Vrel_mean_ii}). The blue-dashed line is the 
non-relativistic Maxwell value $4/\sqrt{\pi x}$. }
\label{Fig1}
\end{figure}

\subsection{Constant cross section, s and p wave scattering}

In many cases the cross section in the low-velocity limit goes as $1/v_r$ and the 
product $\sigma v_r$ can be considered constant and factorize out the 
thermal integral. Both for non-relativistic and relativistic averages we have 
$\langle \sigma v_{\text{rel}} \rangle = \sigma v_r$ because of the normalization of the
probability distributions 
$\int^\infty_0 dv_r F_M (v_r)=\int^1_0 dv_\texttt{r} \mathcal{P}(v_\texttt{r})=1$.

If the cross section is velocity independent, $\sigma=k$, then
\begin{flalign}
\langle \sigma {v}_{\texttt{r}} \rangle_{\sigma=k}=\sigma
\langle {v}_{\texttt{r}} \rangle=\sigma
\frac{4}{x}\frac{K_3(2x)}{K^2_2(x)}.
\label{sigma_cost}
\end{flalign}
The exact expression for the mean value of the relativistic
relative velocity was found in Ref.~\cite{Cannoni1}, see also Appendix~\ref{A1}.
In the range of $x$ between 20 and 40 that is typical for masses and freeze-out 
temperatures of weakly interacting massive particles, the mean relativistic
relative velocity is always smaller than the Maxwell's value $4/\sqrt{\pi x}$
as can be seen in Fig.~\ref{Fig1}.

We consider now the case $n=0$ as the dominant 
contribution to the cross section, the so-called $s$-wave scattering.
The  coefficients for $n=0$ of the expansions (\ref{expansion_rho=0}), 
(\ref{expansion_rho=1}), (\ref{expansion_rho>>1}) 
take a simple form in terms of the modified Bessel function:
\begin{flalign}
&\mathcal{K}^{\rho=0}_{n=0}(x)=\frac{K^2_1(x)}{K^2_2(x)},
\label{s_wave_coeff_0}
\\
&\mathcal{K}^{\rho=1}_{n=0}(x)=\frac{2}{x}\frac{K_1(2x)}{K^2_2(x)},
\label{s_wave_coeff_1}
\\
&\mathcal{K}^{\rho\gg 1}_{n=0}(x)= \rho \frac{K^2_1(\sqrt{\rho} x)}{K^2_2(x)},
\label{s_wave_coeff_mm1}
\end{flalign}
The coefficients for the cases $\rho=0$ and $\rho\gg 1$ follows from Eq.~(\ref{Meijer_K_1^2})
given in the Appendix.

In the case of $p$-wave scattering also 
the term with $n=1$ is important.
The coefficients have the following expressions in terms of Bessel functions:
\begin{flalign}
&\mathcal{K}^{\rho=0}_{n=1}(x)= \frac{1}{2}(1-\frac{K^2_1(x)}{K^2_2(x)}),
\label{p_wave_coeff_0}
\\
&\mathcal{K}^{\rho=1}_{n=1}(x)=\frac{4}{x^2}\frac{K_2(2x)}{K^2_2(x)},
\label{p_wave_coeff_1}
\\
&\mathcal{K}^{\rho\gg 1}_{n=1}(x)=\frac{\rho^2}{2} \frac{K^2_1(\sqrt{\rho} x)+K^2_2(\sqrt{\rho} x)}{K^2_2(x)}.
\label{p_wave_coeff_mm1}
\end{flalign}
The coefficient for the $\rho=0$ follows from Eq.~(\ref{G_-3/2_1/2_3/2}) and that 
for $\rho\gg1$ from Eq.~(\ref{G_0_-5/2_1/2_-1/2}).
These exact coefficients that we found in the case of a constant cross section, $s$ and $p$ wave
scattering can be useful for a rapid estimation of the thermal average.
\begin{figure}
\includegraphics*[scale=.6]{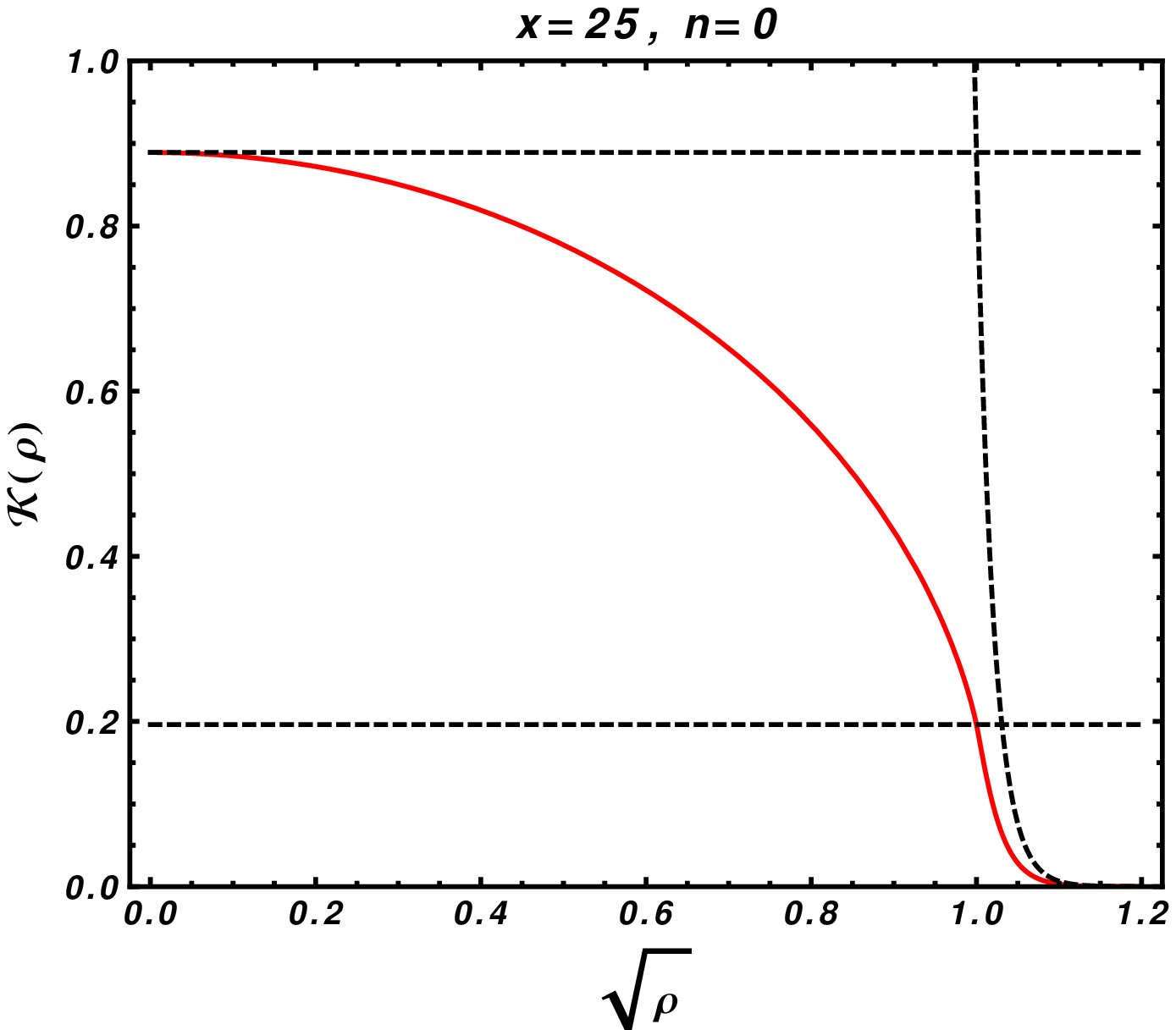}
\includegraphics*[scale=.6]{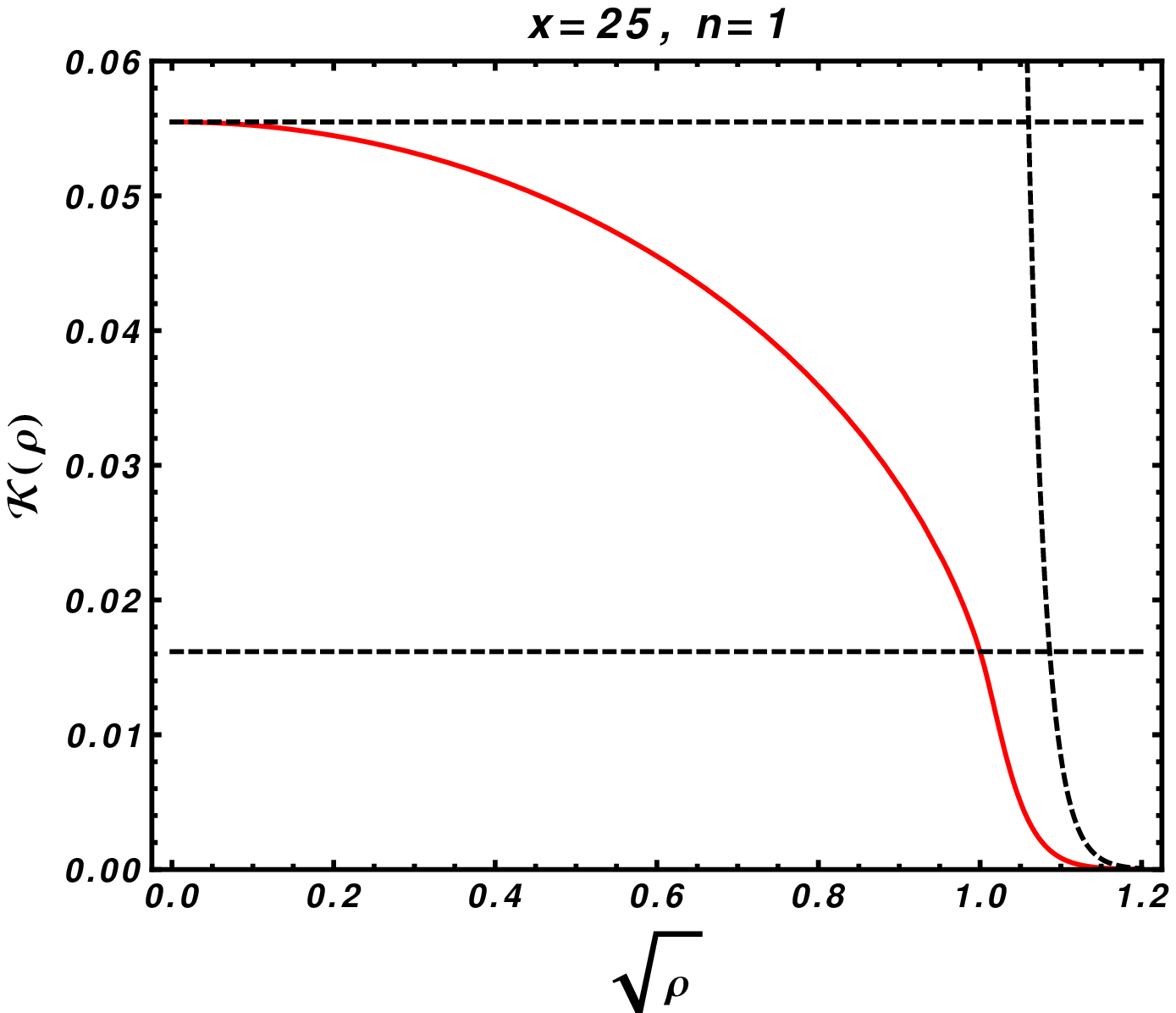}
\caption{Coefficients of the expansion (\ref{expansion}) as a function of the threshold parameter $\sqrt{\rho}=m_f /m$ .
$s$ wave scattering, top panel, and $p$ wave scattering, bottom panel, at $x=25$. 
The red lines are the exact relativistic 
value obtained by numerical integration of (\ref{I_n}). The black-dashed lines correspond 
to the three particular values $\rho=0$, $\rho=1$, $\rho\gg1$ 
given by Eqs.~(\ref{s_wave_coeff_0}), (\ref{s_wave_coeff_1}), (\ref{s_wave_coeff_mm1}) for $n=0$
and Eqs.~(\ref{p_wave_coeff_0}), (\ref{p_wave_coeff_1}), (\ref{p_wave_coeff_mm1}) for $n=1$.  }
\label{Fig2}
\end{figure}  
In Fig.~\ref{Fig2} we plot the coefficients of the expansion (\ref{expansion}) 
as a function of $\sqrt{\rho}=m_{f} /m $ at $x=25$ for the cases $n=0$, top panel, and $n=1$, bottom panel. The red line 
is exact and obtained with numerical integration of (\ref{K_n_rho}) while the dashed lines 
are the constant values given by the formulas (\ref{s_wave_coeff_0}), (\ref{s_wave_coeff_1}), 
(\ref{s_wave_coeff_mm1}) for $n=0$, and (\ref{p_wave_coeff_0}), (\ref{p_wave_coeff_1}), 
(\ref{p_wave_coeff_mm1}) for $n=1$.
The agreement show the correctness of the above formulas.

\section{Summary and final remarks}
\label{summary}

In this paper we have clarified the meaning of $\langle \sigma v_{\text{rel}}\rangle$
in the relativistic framework.
The velocity $v_{\text{rel}}$ is actually the invariant relative velocity 
$V_{\texttt{r}}$ defined by~(\ref{vRelRel}), (\ref{vRel_p1p2}), (\ref{vrel_s}). 
The thermal average of $\sigma V_{\texttt{r}}$ is given by the integral over
probability distribution of $V_{\texttt{r}}$, Eq.~(\ref{P_v}), in the same 
way as the non-relativistic average is determined by the non-relativistic 
relative velocity $v_r$ and the Maxwell distribution. 

We have remarked that the  M\o{}ller velocity is not a fundamental
physical velocity and it is at the origin of the conceptual issues discussed 
in the paper.
Its use as a physical quantity should be avoided
in favor of the true invariant relative velocity $V_{\texttt{r}}$.

We have found that the coefficients of the low-velocity expansion of 
$\langle \sigma V_{\texttt{r}}\rangle$ admit exact analytical representation
in the cases that masses of the final state particles are $m_f =0$, $m_f=m$ and $m_f \gg m$. 
The coefficients are given by the generalized hypergeometric Meijer $G$ functions 
and can be reduced to expressions involving combinations of modified Bessel functions $K_\nu$.

\section*{Acknowledgments}

The author acknowledges N.~Fornengo and M.~Peir\'{o} for useful discussions 
during the 9th MultiDark workshop, Halcal\'{a} de Henares, 6-9 November 2013, Spain,
where some of the results of the present paper and of Ref.~\cite{Cannoni1} were presented. 
Work supported in part by MultiDark under Grant No. CSD2009-00064 of the
Spanish MICINN  Consolider-Ingenio 2010 Program. Further support is provided by the
MICINN project FPA2011-23781 and from the Grant MICINN-INFN(PG21)AIC-D-2011-0724. 

\appendix
\section{$\boldsymbol{\textit{G}}$ functions and integrals of $\boldsymbol{\textit{K}_\nu}$  }
\label{AppendixA}

In this Appendix we show how some integrals involving 
the products of powers and the modified Bessel function $K_1$
can be evaluated in terms of combination of other Bessel functions $K_n$
by using the generalized hypergeometric Meijer's $G$ function
$G^{m,n}_{p,q}(x|\{a_1,...,a_n,...,a_p \},\{b_1,...,b_m,...,b_q\})$. 
Detailed definition and relations with other special and 
usual functions can be found for example in~\cite{GR}.
The same kind of integrals were also found in Ref.~\cite{Cannoni1}.

The $G$ function is usually written  in tabular representation as
\begin{equation}
G^{m,n}_{p,q}\left(z\Bigg\rvert
\begin{array}{c}
 a_1,...,a_n,a_{n+1},...,a_p\\
b_1,...,b_m,b_{m+1}...,b_q
\end{array}
\right).
\end{equation}
Note that the top-left index $m$ counts the first $m$ bottom-left parameters,
the bottom-right index $p$ counts the top-right parameters and so on.
The function is invariant for permutations of the top $a_i$ parameters 
or permutations of the $b_j$'s.

The $G$ functions can be simplified thanks to the property 
\begin{flalign}
G^{m,n}_{p,q}
\left(z\Bigg\rvert
\begin{matrix}{}
a_1-\alpha,...,a_p-\alpha\\
b_1 -\alpha,...,b_q -\alpha
\end{matrix}
\right)
=\frac{1}{z^{\alpha}}
G^{m,n}_{p,q}
\left(z\Bigg\rvert
\begin{matrix}{}
a_1,...,a_p\\
b_1,...,b_q 
\end{matrix}
\right)
\label{Prop_meno_alpha},
\end{flalign}
and  when one of the upper index is equal to one of the 
lower index the function is reduced, for example if $a_p=b_q=c$
\begin{flalign}
G^{m,n}_{p,q}
\left(z\Bigg\rvert
\begin{matrix}{}
a_1,...,c\\
b_1,...,c 
\end{matrix}
\right)
=
G^{m-1,n}_{p-1,q-1}
\left(z\Bigg\rvert
\begin{matrix}{}
a_1,...,a_{p-1}\\
b_1,...,b_{q-1} 
\end{matrix}
\right).
\label{Prop_2_uguali}
\end{flalign}
The relations between the modified Bessel 
functions and the Meijer function are 
\begin{flalign}
%&K_{-\nu}(z)=K_{\nu}(z),
%\label{nu_aprity}
%\\
&G_{0,2}^{2,0}\left(\frac{z^2}{4}\Bigg\rvert
\begin{array}{c}
\\
\frac{\delta+\nu}{2},\frac{\delta -\nu}{2}
\end{array}
\right)=\frac{z^\delta}{2^{\delta-1}} K_\nu (z),
\label{K_2020}
\\
&G_{1,3}^{3,0}\left(x^2\Bigg\rvert
\begin{array}{c}
 0 \\
 -\frac{1}{2},{\nu}+\frac{1}{2},\nu-\frac{1}{2}
\end{array}
\right)
=\frac{2}{\sqrt{\pi}}\frac{K^2_\nu(x)}{ x}.
\label{Meijer_K_1^2}
\end{flalign}
\begin{flalign}
&\int_{1}^{\infty} dz z^{\lambda} (z-1)^{\mu-1} K_\nu (a\sqrt{z})=\nonumber\\
&\;\;\;\;\Gamma(\mu) 2^{2\lambda-1} a^{-2\lambda} G_{1,3}^{3,0}
\begin{pmatrix}
\frac{a^2}{4} \Biggr\rvert
\begin{array}{c}
 0 \\
-\mu, \frac{\nu}{2}+\lambda,-\frac{\nu}{2}+\lambda
\end{array}
\end{pmatrix}.
\label{I_GR}
\end{flalign}

\subsection{Mean relative velocity}
\label{A1}

In this sub-section we explicitly evaluate the mean value of the relative velocity~(\ref{Vrel_mean_ii}).
For the derivation of the  expression with $m_1 \neq m_2$ see Ref.~\cite{Cannoni1}.  
We evaluate the integral $\int_{0}^{\infty} d\gamma_\texttt{r} \mathcal{P}(\gamma_\texttt{r}) v_\texttt{r} $ 
with $V_{\texttt{r}}=\sqrt{\gamma_{\texttt{r}}-1}\sqrt{\gamma_{\texttt{r}}+1}/\gamma_\texttt{r}$. Changing 
variable to $z=(\gamma_\texttt{r}+1)/2$ we find
\begin{equation}
\langle V_{\texttt{r}} \rangle 
=\frac{4 x}{K^2_2(x)}
\int_{1}^{\infty} dz z^{1/2} (z-1) K_1(2x\sqrt{z}).
\end{equation}
The integral corresponds to Eq.~(\ref{I_GR}) with $\lambda=1/2$, $\mu=2$, $\nu=1$ ,
\begin{flalign}
\mathcal{I}=\frac{1}{2x}
G_{1,3}^{3,0}\left(x^2 \Biggr\rvert
 \begin{array}{c}
  0 \\
  -2,1,0
 \end{array}
 \right)=\frac{1}{2x}
G_{0,2}^{2,0}\left(\frac{(2x)^2}{4} \Biggr\rvert
 \begin{array}{c}
\\
-2,1
 \end{array}
 \right) ,
\end{flalign}
where we used the property (\ref{Prop_2_uguali}) for the second equality.
Now we use Eq.~(\ref{K_2020}) with $\delta=-1$ and $\nu=3$ that gives
\begin{flalign}
\mathcal{I}=\frac{1}{2x} \frac{4}{2x} K_3(2x).
\end{flalign}
Finally, multiplying by ${4 x}/{K^2_2(x)}$, 
\begin{flalign}
\langle V_{\texttt{r}} \rangle=\frac{4}{x} \frac{K_3(2x)}{K^2_2(x)}.
\label{Vrel_mean_ii}
\end{flalign}

\subsection{$\boldsymbol{\textit{G}}$ functions for $\boldsymbol{\rho =1}$}
\label{A2}

In the case $\rho=1$ the coefficients of the expansion involve a reducible $G$ function:
\begin{flalign}
G_{1,3}^{3,0}
\begin{pmatrix}
x^2 \Biggr\rvert
\begin{array}{c}
 0 \\
-n-2, -1, 0
\end{array}
\end{pmatrix}
&=
G_{0,2}^{2,0}
\begin{pmatrix}
\frac{(2x)^2}{4} \Biggr\rvert
\begin{array}{c}
 \\
-n-2, -1
\end{array}
\end{pmatrix}\nonumber\\
&=
\frac{2}{x^{3+n}} K_{n+1} (2x)
\end{flalign}
where we have used Eq.~(\ref{Prop_2_uguali}) and then Eq.~(\ref{K_2020}) with $\delta=-3-n$ and $\nu=-n-1$.

\subsection{$\boldsymbol{\textit{G}}$ functions for $\textit{n}=1$ and $\boldsymbol{\rho=0}$, $\boldsymbol{\rho>>1}$ }
\label{A3}

Changing  variable to $y={(1+\gamma_{\texttt{r}})}/2$,
the normalization condition of the probability distribution in the diagonal case $x_1=x_2=x$
defines two integrals
\begin{flalign}
&\frac{2 x}{K^2_2(x)}
\int_{1}^{\infty} dy (2y-1) \sqrt{y-1} K_1(2x\sqrt{y})\crcr
&\;\;\;\;\;\;\;\;\;\;\;\;=\frac{2 x}{K^2_2(x)}(2I_1 -I_0)=1.
\label{Pz}
\end{flalign}
The  integral 
\begin{flalign}
I_0=\frac{\sqrt{\pi}}{4}
G_{1,3}^{3,0}\left(x^2\Bigg\rvert
\begin{array}{c}
0 \\
-\frac{3}{2},-\frac{1}{2},\frac{1}{2}
\end{array}
\right)
=\frac{K^2_1 (x)}{2x},
\label{Int_0}
\end{flalign}
is evaluated thanks to the basic integral (\ref{I_GR}) with
$\lambda=0$, $\mu=3/2$, $\nu=1$ and then the second equality follows 
from Eq.~(\ref{Meijer_K_1^2}).
The integral (\ref{I_GR}) with $\lambda=1$, $\mu=3/2$, $\nu=1$ reads
\begin{flalign}
I_1=\frac{\sqrt{\pi}}{4x^2}
G_{1,3}^{3,0}\left(x^2\Bigg\rvert
\begin{array}{c}
0 \\
-\frac{3}{2},\frac{1}{2},\frac{3}{2}
\end{array}
\right).
\label{Int_1}
\end{flalign}
Combining Eq.~(\ref{Pz}) with Eqs.~(\ref{Int_0}) and  (\ref{Int_1}) we find
\begin{flalign}
G_{1,3}^{3,0}\left(x^2\Bigg\rvert
\begin{array}{c}
0 \\
-\frac{3}{2},\frac{1}{2},\frac{3}{2}
\end{array}
\right)
=\frac{x}{\sqrt{\pi}}(K^2_1(x) +K^2_2 (x)).
\label{G_-3/2_1/2_3/2}
\end{flalign}
Eq.~(\ref{I_GR}) with $\mu=5/2$, $\lambda=0$, $\nu=1$ gives the difference
\begin{flalign}
I_1 -I_0=\frac{3}{8}\sqrt{\pi}  G_{1,3}^{3,0}\left(x^2\Bigg\rvert
\begin{array}{c}
0 \\
-\frac{5}{2},-\frac{1}{2},\frac{1}{2}
\end{array}
\right).
\end{flalign}
Combining (\ref{Int_0}), (\ref{Int_1}) and (\ref{G_-3/2_1/2_3/2}) we then find
\begin{flalign}
G_{1,3}^{3,0}\left(x^2\Bigg\rvert
\begin{array}{c}
0 \\
-\frac{5}{2},-\frac{1}{2},\frac{1}{2}
\end{array}
\right)=\frac{2}{3} \frac{K^2_2(x)-K^2_1(x)}{\sqrt{\pi}x}.
\label{G_0_-5/2_1/2_-1/2}
\end{flalign}

\end{document}